\documentclass[notitlepage,aps,pra,amsmath,amsfonts,amssymb,floatfix,superscriptaddress,10pt,reprint,longbibliography]{revtex4-2}
\usepackage{color}
\usepackage{braket}
\usepackage{mathtools}
\usepackage{dcolumn} 
\usepackage{graphicx}
\usepackage{hyperref}

\begin{document}
\title{Seniority Eigenstate Configuration Interaction}

\author{Thomas M. Henderson}
\email[Author to whom correspondence should be addressed: ]{thomas.henderson@rice.edu}
\affiliation{Department of Chemistry, Rice University, Houston, TX 77005-1892, USA}
\affiliation{Department of Physics and Astronomy, Rice University, Houston, TX 77005-1892, USA}

\author{Guo P. Chen}
\altaffiliation{Current address: Advanced Materials Thrust, The Hong Kong University of Science and Technology (Guangzhou), Guangdong 511400, China}
\affiliation{Department of Chemistry, Rice University, Houston, TX 77005-1892, USA}

\author{Gustavo E. Scuseria}
\affiliation{Department of Chemistry, Rice University, Houston, TX 77005-1892, USA}
\affiliation{Department of Physics and Astronomy, Rice University, Houston, TX 77005-1892, USA}
\date{\today}

\begin{abstract}
Zero-seniority methods have shown great promise for the description of strongly-correlated electronic systems.  Other seniority sectors have been much less explored, and in particular the maximal seniority sector and zero seniority have the same underlying algebraic structure.  We introduce a seniority eigenstate configuration interaction in which the wave function is constrained to have good fixed local seniority for each paired orbital, by which we mean we partition orbitals into a pairing set with seniority zero, and a spin set with seniority one.  We show how to build the effective Hamiltonian for this ansatz, and demonstrate that high-seniority wave functions have unexpectedly excellent accuracy for strongly-correlated fermionic systems, with accuracy competitive with or better than seniority zero for the Hubbard model and for the dissociation of the nitrogen molecule.
\end{abstract}

\maketitle

\section{Introduction}
For weakly-correlated electronic systems, we frequently organize Hilbert space by excitation level.  The idea is simple.  We begin by choosing a suitable set of spinorbitals.  We divide them into occupied spinorbitals and virtual spinorbitals, and then we can determine the excitation level from the number of virtual spinorbitals that contain electrons and the number of occupied spinorbitals that do not.  Excitation level is a property of a given Slater determinant rather than of a correlated wave function; correlated wave functions generally consist of many different excitation levels because excitation level is not a symmetry of the Hamiltonian.  The utility of the excitation level concept lies in the observation that for a weakly-correlated system with suitably chosen orbitals and occupied-virtual partitioning schemes, the important dynamic correlations are given by a wave function that has low excitation level or are constructed by operators (e.g. the cluster operators of coupled cluster theory) that increase excitation level only to a small degree.

In the past few decades, seniority has emerged as a similarly useful organizing principle for the description of strongly-correlated fermionic systems \cite{Bytautas2011,Johnson2012,Alcoba2013,Alcoba2014,Alcoba2014a,Chen2015,Bytautas2015,Kossoski2022,Kossoski2023,CaleroOsorio2025}.  As with excitation level, seniority is orbital dependent, and we must make a further choice: where excitation level requires us to partition the spinorbitals into occupied and virtual levels, seniority requires us to choose, for every spinorbital $P$, a paired counterpart $\bar{P}$.  The seniority $\Omega_p$ of the paired level $p = \{P,\bar{P}\}$ is 1 if $p$ is singly-occupied and $0$ if the level is doubly-occupied or empty; the seniority $\Omega$ of a determinant is just the sum of the seniorities of the paired levels.  Seniority is not generally a symmetry of the Hamiltonian so correlated wave functions consist of many different seniorities, but we observe that with a suitably chosen set of spinorbitals and a suitable pairing scheme, very often the most important strong correlations can be described with wave functions that have low seniority.

The seniority zero sector, where no level is singly-occupied, constitutes a paired fermionic system with an underlying $\mathfrak{su}(2)$ algebra for each level, and such wave functions have a particularly appealing mathematical structure.  The exact result within this sector is referred to as doubly-occupied configuration interaction \cite{Allen1962, Smith1965, Weinhold1967, Veillard1967,Couty1997,Kollmar2003,Bytautas2011,Alcoba2024} (DOCI) and is reasonably approximated by the antisymmetrized product of one-reference orbital geminals \cite{Limacher2013,Boguslawski2014}, also known as pair coupled cluster doubles \cite{Stein2014,Henderson2014b}.

Other seniority sectors are less explored, but the \textit{maximal} seniority sector, where every level is singly occupied, has the same mathematical structure as seniority zero \cite{MartinezGonzalez2025}, with an underlying $\mathfrak{su}(2)$ algebra that in this case gives us what is, in practice, a spin 1/2 system.  All of the mathematical machinery developed for seniority zero can be trivially extended to study maximal seniority by use of a Nambu transformation (see below) \cite{Nambu1960}, the basic idea of which is to map $c_P^\dagger$ to $a_P^\dagger$ but $c_{\bar{P}}^\dagger$ to $a_{\bar{P}}$.  That is, we replace occupied ``barred'' levels $\bar{P}$ with empty and vice versa, so that if we have occupied $P$ but not $\bar{P}$, after transformation we have occupied both; if we have occupied $\bar{P}$ but not $P$, after transformation we have occupied neither.

Maximal seniority has an important limitation which renders it difficult to apply to a typical fermionic system: because each pair of levels $\{P,\bar{P}\}$ is singly occupied, the maximal seniority state is always half-filled.  In practice, this means we must consider states of intermediate seniority, in which some levels are seniority zero and some levels are seniority one.  It is this problem we wish to tackle here.

The intermediate seniority problem is computationally cumbersome in general.  We thus make a simplifying assumption: we will consider states with fixed local seniority, i.e. every determinant in our expansion will have the same $\Omega_p$ for a given pair level $p$.  One might envision this as an active space.  For example, in the dissociation of N$_2$ (considered below), it seems reasonable to suggest that the dissociation limit should place three unpaired electrons on each of the two nitrogen atoms, so we should work with local seniority six, corresponding basically to six spatial orbitals formed from the $2p$ atomic orbitals that are singly occupied, with the remaining spatial orbitals to be doubly occupied or empty.

As we shall see, this simple idea leads to surprisingly reasonable results for strong correlations in both the Hubbard model \cite{Hubbard1963} and N$_2$ dissociation.  Weak correlations must still be included, and serve both to couple seniority sectors and, in general, to break local seniority even when global seniority is conserved.  We defer consideration of this problem to a later date.


\section{The Seniority-Conserving Hamiltonian}
We find it most convenient to frame everything in terms of an effective $\mathfrak{so}(4)$ Hamiltonian that conserves the seniority of each single-particle level.  In previous work we have presented the closed shell case \cite{Henderson2015}; here, we show the spin-unrestricted extension.  One could also do a spin general extension, but we have not considered it here.

Suppose we have a two-body Hamiltonian written in an unrestricted spinorbital basis in which each spinorbital has definite $S_z$.  Such a Hamiltonian is of the form
\begin{align}
H &= E_0 + \sum_{pq} \sum_{\sigma \in \{\uparrow,\downarrow\}} h_{pq}^\sigma \, c_{p_\sigma}^\dagger \, c_{q_\sigma}
\\
  &+ \frac{1}{2} \, \sum_{pqrs} \sum_{\sigma \in \{\uparrow,\downarrow\}} v_{pqrs}^{\sigma\sigma} \, c_{p_\sigma}^\dagger \, c_{q_\sigma}^\dagger \, c_{s_\sigma} \, c_{r_\sigma}
\nonumber
\\
 &+ \sum_{pqrs} v_{pqrs}^{\uparrow\downarrow} \, c_{p_\uparrow}^\dagger \, c_{q_\downarrow}^\dagger \, c_{s_\downarrow} \, c_{r_\uparrow},
\nonumber
\end{align}
where $v_{pqrs}$ is a two-electron integral in Dirac order and is not antisymmetrized.

In a manner akin to what we have already discussed, we can extract a (local) seniority-conserving portion of this Hamiltonian in which we pair levels $p_\uparrow$ and $p_\downarrow$ by restricting summation indices.  In the first term we require $p=q$.  In the second term we require $p \ne q$ (because $c_{p_\sigma}^\dagger \, c_{p_\sigma}^\dagger = 0$) and can either have $pq=rs$ or $pq=sr$.  In the third term we can have $p=q$ and $r=s$, or, with $p \ne q$, we can have $pq=rs$ or $pq=sr$; we have excluded $p=q$ in this last set of possibilities because it is already covered.

We will omit the derivation here for brevity (it can be found in the Supporting Information), but the result is that, after identifying
\begin{subequations}
\begin{align}
c_{p_\uparrow}^\dagger \, c_{p_\uparrow} &= n_{p_\uparrow} = \frac{1}{2} \, N_p + S_p^z,
\\
c_{p_\downarrow}^\dagger \, c_{p_\downarrow} &= n_{p_\downarrow} = \frac{1}{2} \, N_p - S_p^z,
\\
c_{p_\uparrow}^\dagger \, c_{p_\downarrow} &= S_p^+,
\\
c_{p_\downarrow}^\dagger \, c_{p_\uparrow} &= S_p^-,
\\
c_{p_\uparrow}^\dagger \, c_{p_\downarrow}^\dagger &= P_p^\dagger,
\\
c_{p_\downarrow} \, c_{p_\uparrow} &= P_p,
\end{align}
\end{subequations}
we can identify the local seniority-conserving Hamiltonian as
\begin{align}
H_{\delta\Omega=0}
 &= E_0
\label{Eqn:Hamiltonian}
\\
 &+ \sum_p \epsilon_p \, N_p
  + \sum_{pq} L_{pq} \, P_p^\dagger \, P_q
  + \frac{1}{4} \, \sum_{p \ne q} W_{pq} \, N_p \, N_q
\nonumber
\\
 &+ \sum_p B_p \, S_p^z
  - \sum_{p \ne q} K_{pq}^{\uparrow\downarrow} \, S_p^+ \, S_q^-
  + \sum_{p \ne q } B_{pq} \, S_p^z \, S_q^z
\nonumber
\\
 &+ \sum_{p \ne q} X_{pq} \, N_p \, S_q^z.
\nonumber
\end{align}
Here, the various Hamiltonian parameters are
\begin{subequations}
\begin{align}
\epsilon_p &= \frac{1}{2} \, \left(h_{pp}^\uparrow + h_{pp}^\downarrow\right),
\\
L_{pq} &= v_{ppqq}^{\uparrow\downarrow},
\\
W_{pq} &= \frac{1}{2} \, J_{pq}^{\uparrow\uparrow} + \frac{1}{2} \, J_{pq}^{\downarrow\downarrow} + J_{pq}^{\uparrow\downarrow} - \frac{1}{2} \, K_{pq}^{\uparrow\uparrow} - \frac{1}{2} \, K_{pq}^{\downarrow\downarrow},
\\
B_p &= h_{pp}^\uparrow - h_{pp}^\downarrow,
\\
B_{pq} &= \frac{1}{2} \, J_{pq}^{\uparrow\uparrow} + \frac{1}{2} \, J_{pq}^{\downarrow\downarrow} - J_{pq}^{\uparrow\downarrow} - \frac{1}{2} \, K_{pq}^{\uparrow\uparrow} - \frac{1}{2} \, K_{pq}^{\downarrow\downarrow},
\\
X_{pq} &= \frac{1}{2} \, \left(J_{pq}^{\uparrow\uparrow} - J_{pq}^{\downarrow\downarrow}\right) - \frac{1}{2} \, \left(K_{pq}^{\uparrow\uparrow} - K_{pq}^{\downarrow\downarrow}\right)
\\
 & - \frac{1}{2} \, \left(J_{pq}^{\uparrow\downarrow} - J_{qp}^{\uparrow\downarrow}\right)
\nonumber
\end{align}
\end{subequations}
in terms of the Coulomb integrals
\begin{equation}
J_{pq}^{\sigma\sigma^\prime} = v_{pqpq}^{\sigma\sigma^\prime} = \bra{p_\sigma q_{\sigma^\prime}}v\ket{p_\sigma q_{\sigma^\prime}}
\end{equation}
and exchange integrals
\begin{equation}
K_{pq}^{\sigma\sigma^\prime} = v_{pqqp}^{\sigma\sigma^\prime} = \bra{p_\sigma q_{\sigma^\prime}}v\ket{q_\sigma p_{\sigma^\prime}}.
\end{equation}

This Hamiltonian conserves the seniority of every level $p$.  A level is either a pairing level (doubly occupied or empty) or a spin level (singly occupied).  The $L_{pq}$ term interconverts empty and doubly-occupied levels, while the $K_{pq}^{\uparrow\downarrow}$ term interconverts levels with $\uparrow$ spin and levels with $\downarrow$ spin.  The other terms, depending only on $N_p$ and $S_p^z$, have determinants as eigenstates.  We emphasize that this effective Hamiltonian is just all the local-seniority--conserving terms in the original electronic Hamiltonian, written in terms of the $\mathfrak{so}(4)$ operators for algebraic convenience.

\section{Seniority Eigenstate Configuration Interaction}
Because the Hamiltonian of Eqn. \ref{Eqn:Hamiltonian} conserves the seniority of every level, we can label each of its eigenstates by the seniority of each orbital, partitioning the orbitals into two subsets.  The first, with orbitals denoted by indices $\lambda$, $\mu$, $\nu$, \ldots, are seniority zero.  These are never singly occupied, and are annihilated by $S_\mu^\pm$ and by $S_\mu^z$.  The second, denoted by indices $\alpha$, $\beta$, $\gamma$, \ldots, are always singly occupied; they are annihilated by $P_\alpha^\dagger$ and $P_\alpha$ and are eigenstates of $N_\alpha$ with eigenvalue 1.

Having made this partitioning, we can write an effective Hamiltonian for a given orbital partitioning.  It is
\begin{align}
H_{\delta\Omega=0}^\mathrm{eff}
 &=  \left(E_0 + \sum_\alpha \epsilon_\alpha + \frac{1}{4} \, \sum_{\alpha \ne \beta} W_{\alpha\beta}\right)
\label{Eqn:HSECI}
\\
 &+ \sum_\mu \left(\epsilon_\mu + \frac{1}{2} \, \sum_\alpha W_{\mu\alpha}\right) \, N_\mu
\nonumber
\\
 &+ \sum_{\mu\nu} L_{\mu\nu} \, P_\mu^\dagger \, P_\nu + \frac{1}{4} \, \sum_{\mu \ne \nu} W_{\mu\nu} \, N_\mu \, N_\nu
\nonumber
\\
 &+ \sum_\alpha \left(B_\alpha + \sum_{\beta \ne \alpha} X_{\beta\alpha}\right) \, S_\alpha^z
\nonumber
\\
 &- \sum_{\alpha \ne \beta} K_{\alpha\beta}^{\uparrow\downarrow} \, S_\alpha^+ \, S_\beta^-
  + \sum_{\alpha \ne \beta} B_{\alpha\beta} \, S_\alpha^z \, S_\beta^z
\nonumber
\\
 &+ \sum_{\mu\alpha} X_{\mu\alpha} \, N_\mu \, S_\alpha^z.
\nonumber
\end{align}
One part of this Hamiltonian acts only on the zero-seniority levels, another only on the seniority one levels, and there is a coupling term $\sum_{\mu \alpha} X_{\mu\alpha} \, N_\mu \, S_\alpha^z$.  If we use indices $\Lambda$ for product states $\ket{\Phi_\Lambda}$ in the zero-seniority levels and $\Gamma$ for product states $\ket{\Xi_\Gamma}$ in the seniority one levels, the eigenstates of this effective Hamiltonian are
\begin{equation}
\ket{\Psi} = \sum_{\Gamma\Lambda} C_{\Gamma\Lambda} \, \ket{\Xi_\Gamma} \otimes \ket{\Phi_\Lambda}.
\label{Eqn:SECIState}
\end{equation}
We refer to wave functions of this type with optimized coefficients $C_{\Gamma\Lambda}$ as ``seniority eigenstate configuration interaction'' states, or SECI states for short.  It includes both standard DOCI (where there are no singly-occupied levels) and a spin analog (where every level is singly occupied) as special cases.

For pairing DOCI (i.e. DOCI at seniority zero), there are no levels with indices $\alpha$ or $\beta$, so the effective Hamiltonian reduces to
\begin{align}
H_{\delta\Omega=0}^\mathrm{eff} = E_0 &+ \sum_\mu \epsilon_\mu \, N_\mu + \sum_{\mu\nu} L_{\mu\nu} \, P_\mu^\dagger \, P_\nu
\\
 &+ \frac{1}{4} \, \sum_{\mu \ne \nu} W_{\mu\nu} \, N_\mu \, N_\nu.
\nonumber
\end{align}
For spin DOCI (DOCI at maximal seniority), there are no levels with indices $\mu$ or $\nu$, so the effective Hamiltonian becomes
\begin{align}
H_{\delta\Omega=0}^\mathrm{eff}
 &= \left(E_0 + \sum_\alpha \epsilon_\alpha + \frac{1}{4} \, \sum_{\alpha \ne \beta} W_{\alpha\beta}\right)
\\
 &+ \sum_\alpha \left(B_\alpha + \sum_{\beta \ne \alpha} X_{\beta\alpha}\right) \, S_\alpha^z
\nonumber
\\
 &- \sum_{\alpha \ne \beta} K_{\alpha\beta}^{\uparrow\downarrow} \, S_\alpha^+ \, S_\beta^- + \sum_{\alpha \ne \beta} B_{\alpha\beta} \, S_\alpha^z \, S_\beta^z.
\nonumber
\end{align}

Finally, the Hamiltonian we have written is in an unrestricted spinorbital basis, so that $\uparrow$-spin and $\downarrow$-spin orbitals are permitted to have different spatial components.  If the orbitals are restricted, then $\boldsymbol{J}^{\uparrow\uparrow} = \boldsymbol{J}^{\uparrow\downarrow} = \boldsymbol{J}^{\downarrow\downarrow} = \boldsymbol{J}$, and similarly for $\boldsymbol{K}$.  Then we have simply $W_{pq} = 2 \, J_{pq} - K_{pq}$, $B_p = 0$, $B_{pq} = -K_{pq}$, and $X_{pq} = 0$.  With these simplifications, the effective Hamiltonian adopts a simpler form
\begin{align}
H_{\delta\Omega=0}
 &= E_0
  + \sum_p \epsilon_p \, N_p
  + \sum_{pq} L_{pq} \, P_p^\dagger \, P_q
\\
 &+ \frac{1}{4} \, \sum_{p \ne q} W_{pq} \, N_p \, N_q
  - \sum_{p \ne q} K_{pq} \, \vec{S}_p \cdot \vec{S}_q.
\nonumber
\end{align}
If one then repeats the partitioning into spin and pairing levels, we see that the two problems become completely decoupled:
\begin{align}
H_{\delta\Omega=0}^\mathrm{eff}
 &= \left(E_0 + \sum_\alpha \epsilon_\alpha + \frac{1}{4} \, \sum_{\alpha\ne\beta} W_{\alpha\beta}\right)
\\
 &+ \sum_\mu \left(\epsilon_\mu + \frac{1}{2} \, \sum_\alpha W_{\mu\alpha}\right) \, N_\mu
\nonumber
\\
 &+ \sum_{\mu\nu} L_{\mu\nu} \, P_\mu^\dagger \, P_\nu + \frac{1}{4} \, \sum_{\mu \ne \nu} W_{\mu\nu} \, N_\mu \, N_\nu
\nonumber
\\
 &- \sum_{\alpha \ne \beta} K_{\alpha\beta} \, \vec{S}_\alpha \cdot \vec{S}_\beta.
\nonumber
\end{align}
In this case, the SECI wave function just becomes the direct product of pairing DOCI in the pairing levels and spin DOCI in the spin levels.  This has an appealingly simple structure, but as we shall see, is \textit{too} simple.

\subsection{The Need for Unrestricted Orbitals}
Consider a two-site Hubbard model
\begin{equation}
H = -t \, \sum_\sigma \left(c_{1_\sigma}^\dagger \, c_{2_\sigma} + h.c.\right) + U \, \sum_p n_{p_\uparrow} \, n_{p_\downarrow}.
\end{equation}
For large $U$, it should make sense to treat the problem as seniority two (indeed, as $U$ tends to infinity, the ground state occupies each lattice site once, coupling together to give a singlet, and this state is a seniority eigenstate).  If we do so, and use restricted orbitals, we arrive at
\begin{equation}
H_{\delta\Omega=0}^\mathrm{eff} = E_0 + \sum_p \epsilon_p + \frac{1}{2} \, W_{12} - 2 \, K_{12} \, \vec{S}_1 \cdot \vec{S}_2.
\end{equation}

Now, $\sum_p \epsilon_p$ is the trace of the one-electron part of the Hamiltonian, which for the Hubbard model vanishes no matter the orbital rotation.  Similarly, $E_0 = 0$.  And because of the local structure of the two-electron integrals, $J_{pq} = K_{pq}$.  That means we reduce to
\begin{equation}
H_{\delta\Omega=0}^\mathrm{eff} = J_{12} \, \left(\frac{1}{2} - 2 \, \vec{S}_1 \cdot \vec{S}_2\right).
\end{equation}

We have two $S_z = 0$ configurations, $\ket{\uparrow\downarrow}$ and $\ket{\downarrow\uparrow}$.  The Hamiltonian matrix becomes
\begin{equation}
H = J_{12} \, \begin{pmatrix} 1 & -1 \\ -1 & 1 \end{pmatrix}
\end{equation}
which has eigenvalues $0$ and $2 \, J_{12}$.  Since the Hubbard interaction is positive, $J_{12} > 0$, and the ground state of this Hamiltonian has zero energy, regardless of $U$.  This is just the energy of the two-site model as $U \to \infty$, but is wrong for finite $U$.  Naturally the $S_z = \pm 1$ sectors fare no better; the interaction vanishes for same-spin electrons, and the on-site term is still zero.

The take-home message is simple: if we consider the maximal seniority case here, with restricted orbitals, we would predict that the Hubbard model for two sites has $E=0$ for all (positive) values of $U$.  This is of course wrong, so even for the simple 2-site Hubbard model we need open-shell orbitals in the spin block.  This result generalizes: for half-filled Hubbard models, independent of lattice and independent of (positive) $U$, using restricted orbitals for the maximal seniority case gives zero energy.  Physically, this is essentially because the one-electron hopping term, which gives us negative energy, appears only in the trace where it vanishes for any set of orbitals, and what we are left with is the purely repulsive two-electron interaction which we minimize by making it vanish.

Now the Hubbard model is not the only problem we envision solving, of course, but we take this result as a kind of ``no go'' argument: the open-shell orbitals \textit{must} be treated spin unrestricted.  

On the other hand, we would like to limit spin symmetry breaking.  For this reason, unless otherwise specified, we use closed-shell restricted orbitals for the pairing part, and open-shell unrestricted orbitals only for the spin part.  This means that SECI includes, at zero seniority, restricted DOCI (RDOCI) and at maximal seniority, unrestricted DOCI (UDOCI) of the spin Hamiltonian (which, see below, we can reframe as standard UDOCI for zero seniority on a transformed Hamiltonian).  When we need to be more specific, we will refer to ``RSECI'' to mean restricted orbitals in both blocks, ``USECI'' to mean unrestricted orbitals in both blocks, and ``RUSECI'' to mean restricted orbitals in the pairing block and unrestricted orbitals in the spin block; ``SECI'' without qualification thus refers to RUSECI.

\subsection{Nambu Transformation}
We can interconvert pairing and spin channels by using the Nambu transformation \cite{Nambu1960}
\begin{subequations}
\begin{align}
c_{p_\uparrow}^\dagger &\mapsto a_{p_\uparrow}^\dagger,
\\
c_{p_\downarrow}^\dagger &\mapsto a_{p_\downarrow}.
\end{align}
\end{subequations}
This gives us
\begin{subequations}
\begin{align}
c_{p_\uparrow}^\dagger \, c_{p_\uparrow} &\mapsto a_{p_\uparrow}^\dagger \, a_{p_\uparrow},
\\
c_{p_\downarrow}^\dagger \, c_{p_\downarrow} &\mapsto a_{p_\downarrow} \, a_{p_\downarrow}^\dagger = 1 - a_{p_\downarrow}^\dagger \, a_{p_\downarrow},
\\
c_{p_\uparrow}^\dagger \, c_{p_\downarrow} &\mapsto a_{p_\uparrow}^\dagger \, a_{p_\downarrow}^\dagger,
\\
c_{p_\downarrow}^\dagger \, c_{p_\uparrow} &\mapsto a_{p_\downarrow} \, a_{p_\uparrow},
\\
c_{p_\uparrow}^\dagger \, c_{p_\downarrow}^\dagger &\mapsto a_{p_\uparrow}^\dagger \, a_{p_\downarrow},
\\
c_{p_\downarrow} \, c_{p_\uparrow} &\mapsto a_{p_\downarrow}^\dagger \, a_{p_\uparrow}.
\end{align}
\end{subequations}
Interpreted in terms of the $\mathfrak{so}(4)$ objects $\{P_p,P_p^\dagger,N_p,S_p^+,S_p^-,S_p^z\}$ built from $c_p^\dagger$ and $c_p$ on the one hand, and their analogues $\{\mathcal{P}_p,\mathcal{P}_p^\dagger,\mathcal{N}_p,\mathcal{S}_p^+,\mathcal{S}_p^-,\mathcal{S}_p^z\}$ built from $a_p^\dagger$ and $a_p$ on the other, we have
\begin{subequations}
\begin{align}
P_p &\mapsto \mathcal{S}_p^-,
\\
P_p^\dagger &\mapsto \mathcal{S}_p^+,
\\
N_p &\mapsto 2 \, \mathcal{S}_p^z + 1,
\\
S_p^+ &\mapsto \mathcal{P}_p^\dagger,
\\
S_p^- &\mapsto \mathcal{P}_p,
\\
S_p^z &\mapsto \frac{1}{2} \, \left(\mathcal{N}_p-1\right).
\end{align}
\end{subequations}

The point we wish to communicate here is simply this: one can always interconvert zero seniority, characterized by the operators $P$, $P^\dagger$, and $N$, with maximal seniority characterized by the operators $S^-$, $S^+$, and $S^z$.  

There is, however, a very important distinction we must emphasize.  For zero seniority, we may have any number of pairs we like, but we are restricted to the global $S^z = 0$ sector of Hilbert space (and in fact, if we use restricted orbitals we are restricted to singlets).  Maximal seniority, on the other hand, may be any $S^z$ we like but is restricted to work at half filling because each level is singly occupied.

\begin{figure*}
\includegraphics[width=\columnwidth]{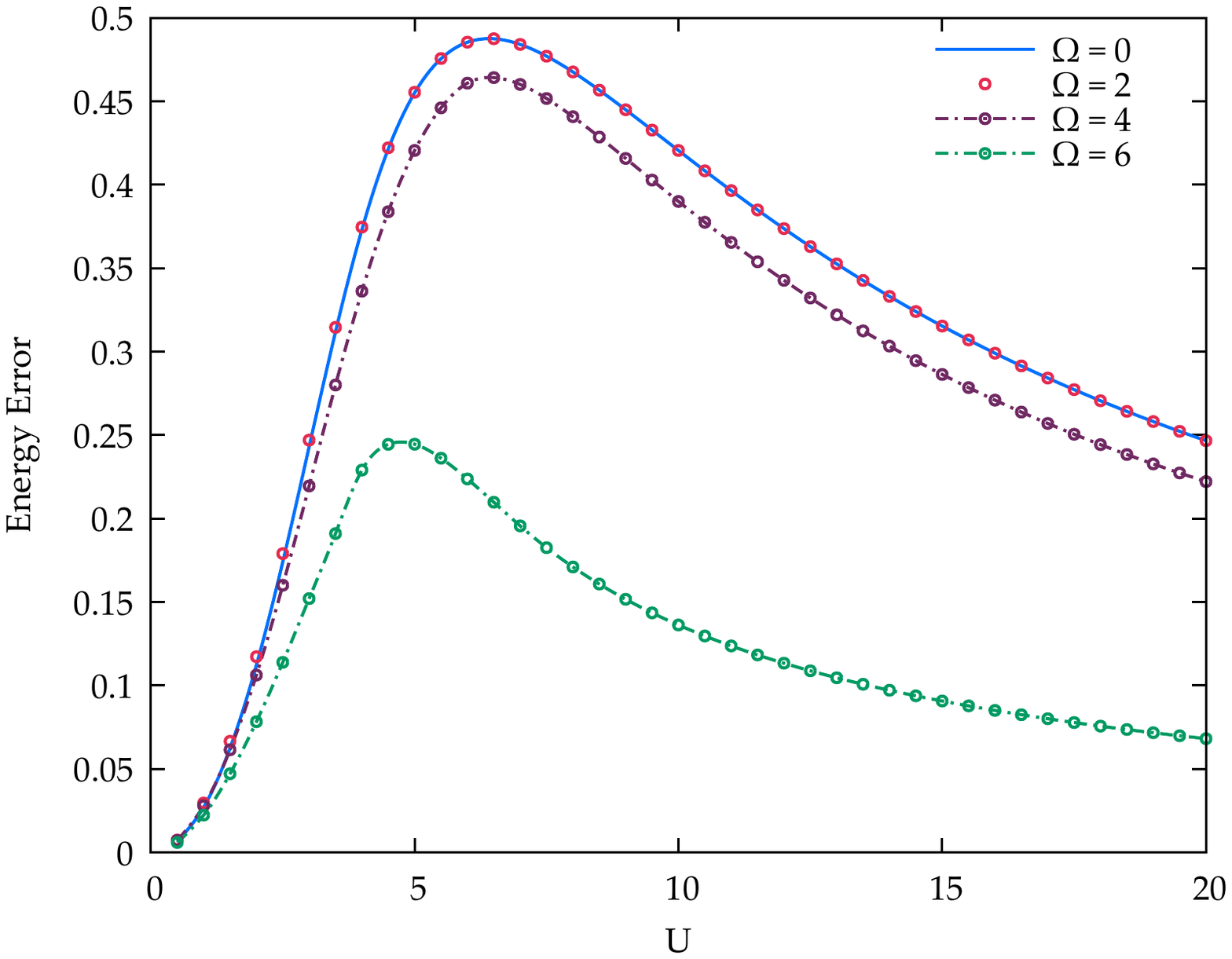}
\hfill
\includegraphics[width=\columnwidth]{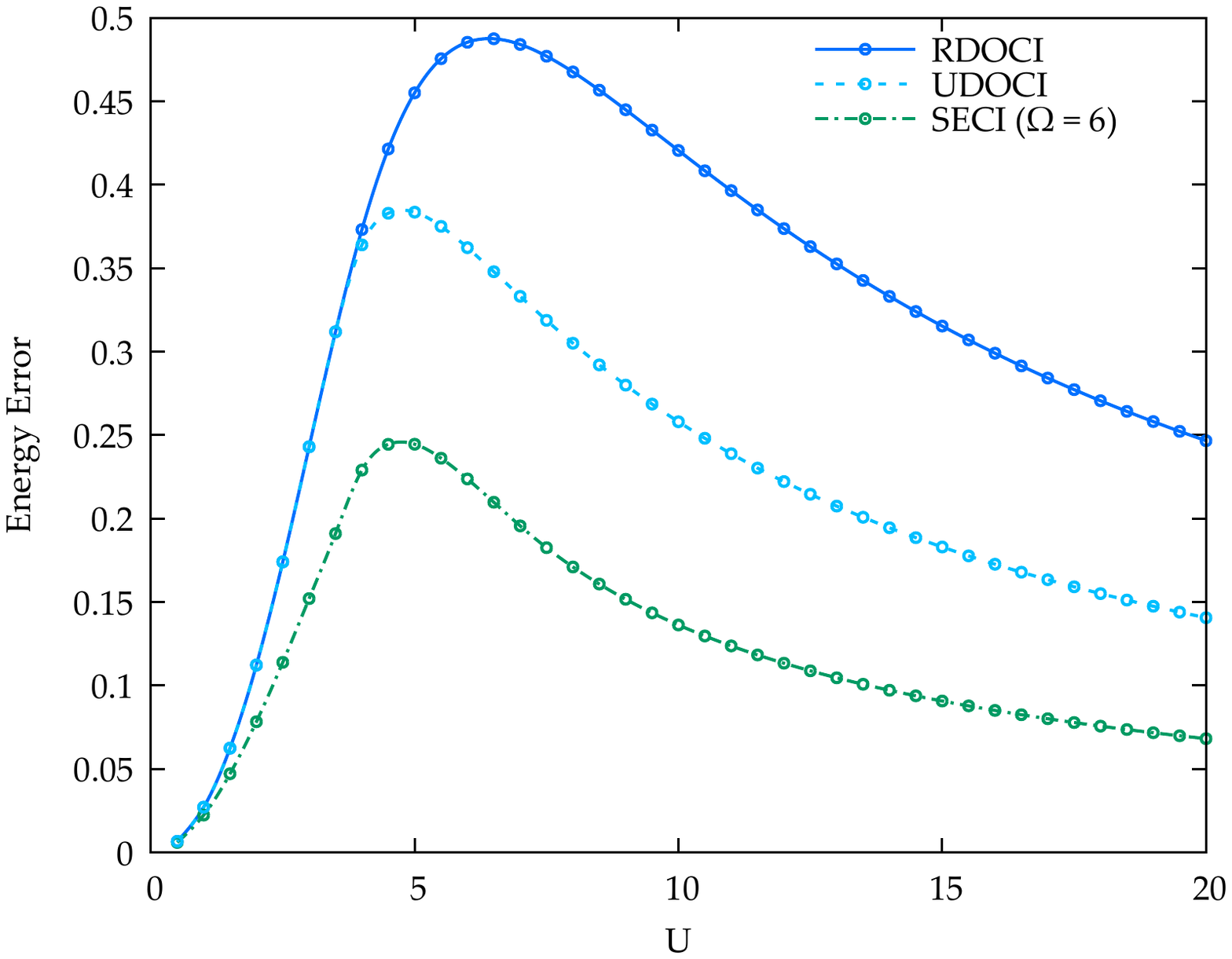}
\caption{Energy errors vs FCI in the 6-site Hubbard model with periodic boundary conditions.  Left panel: Energies for SECI with different seniorities $\Omega$.  Right panel: a comparison of RDOCI (SECI with $\Omega = 0$), UDOCI, and SECI at maximal seniority ($\Omega = 6$).
\label{Fig:Hubbard6}}
\end{figure*}

\section{Results}
We have generated all results with an in-house code.  We optimize the orbitals and coefficients for SECI together, as one does for DOCI.  To initialize the orbital optimization, we use localized Hartree-Fock orbitals; generally we use a UHF calculation so that we can get orbitals in the singly-occupied space.

For the orbital optimization, we write the orbital coefficients $\mathbf{C}_\uparrow$ and $\mathbf{C}_\downarrow$ that take us from the bare Hamiltonian (atomic orbital) basis to the optimized basis as matrix exponentials of antihermitian matrices:
\begin{subequations}
\begin{align}
\mathbf{C}_\uparrow &= \mathrm{e}^{\mathbf{X}},
\\
\mathbf{C}_\downarrow &= \mathrm{e}^{\mathbf{X}} \, \mathrm{e}^\mathbf{Y}
\end{align}
\end{subequations}
where $\mathbf{X}$ is $M \times M$ when we have $M$ spatial orbitals, and $\mathbf{Y}$ is zero except for a $k \times k$ block where $k$ is the number of seniority one levels.  For unrestricted pairing, we simply pair the $p^\mathrm{th}$ $\uparrow$-spin level with the $p^\mathrm{th}$ $\downarrow$-spin level.  We permit the optimization of $\mathbf{X}$ and $\mathbf{Y}$ to dictate the final pairing scheme, and use conjugate gradients to do the orbital optimization.  When the  atomic orbitals are non-orthogonal (as in N$_2$), we choose $\mathbf{C}_\sigma$ as the coefficients that take us from the canonical restricted Hartree-Fock (RHF) orbitals to the optimized orbitals, and let the RHF take care of the non-unitarity.

\subsection{The Importance of High Seniority}
Before we turn to states of intermediate seniority, we wish to establish a point that has been, to the best of our knowledge, almost entirely overlooked: while DOCI (zero seniority, in other words) provides excellent results for strongly-correlated systems, the same is true for maximal seniority eigenstates.

To begin with, then, we start with the 6-site Hubbard model with periodic boundary conditions (PBC), shown in Fig. \ref{Fig:Hubbard6}. The left panel shows the total energy errors for different seniority sectors, from $\Omega = 0$ (which is just restricted DOCI, or RDOCI) through to $\Omega = 6$ (which is the maximal seniority case).  There is no meaningful difference between RDOCI and SECI with $\Omega = 2$, though the results are not numerically identical (they differ by at most 0.005 $t$, which cannot be seen on the plot).  Seniority 4 is better, but the maximal seniority is clearly the best result throughout the curve.  We expected this for large $U/t$ where the lattice sites themselves should be singly occupied.  We were very surprised to see that even at small $U/t$ we still obtain better agreement with FCI by using maximal seniority rather than seniority zero.  Note that the $\Omega = 0$ and $\Omega = 6$ Hilbert spaces are 20-dimensional, while the intermediate seniority spaces are slightly smaller in this case.  For the $\Omega = 2$ and $\Omega = 4$, the Hilbert spaces are of dimension $\tbinom{2}{1} \, \tbinom{4}{2} = 12$.

One might suspect that the significant improvement upon RDOCI for large $U/t$ is an artifact of the difference between restricted and unrestricted orbitals.  To show that this is not obviously the case, we include the right panel of Fig. \ref{Fig:Hubbard6}.  Here, we show that maximal seniority with unrestricted orbitals is also better than unrestricted DOCI (UDOCI), i.e. zero seniority with unrestricted orbitals.  That RDOCI and UDOCI are identical for small $U/t$ should be expected, and clearly SECI has already improved very noticeably upon DOCI before there is even a distinction between RDOCI and UDOCI.

\begin{figure*}
\includegraphics[width=\columnwidth]{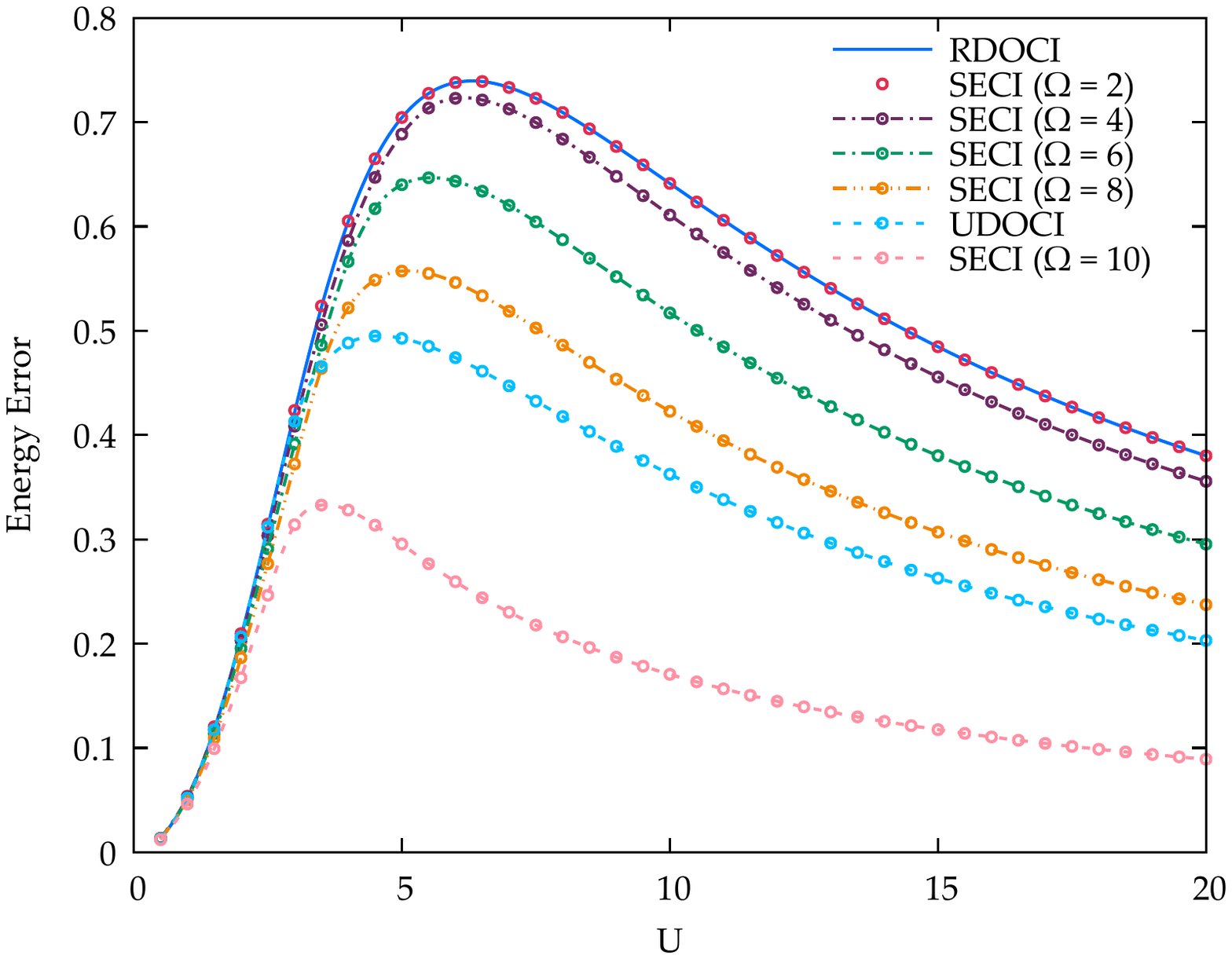}
\hfill
\includegraphics[width=\columnwidth]{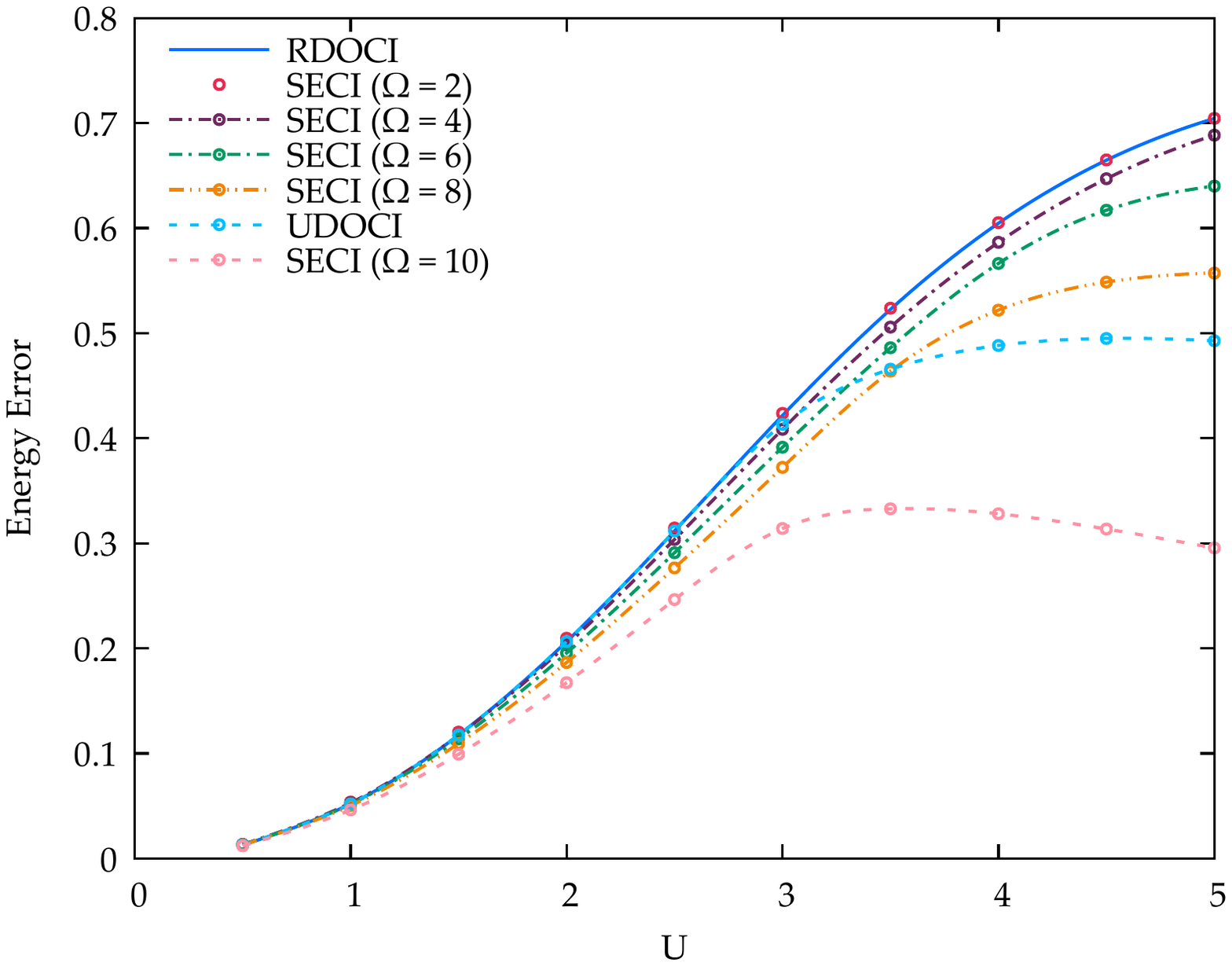}
\caption{Energy errors vs FCI in the 10-site Hubbard model with periodic boundary conditions.  Left panel: Energies for SECI with different seniorities $\Omega$.  Right panel: a zoom in for small $U/t$ to show that even in this regime, SECI with maximal seniority is better than either RDOCI or UDOCI.
\label{Fig:Hubbard10}}
\end{figure*}

We do not wish to belabor this point, but we include similar results for the 10-site half-filled Hubbard model in Fig. \ref{Fig:Hubbard10}.  Again, the improvement in the energy is monotonic: as we increase seniority, we decrease the error in the half-filled Hubbard model, even for small $U/t$, and SECI at maximal seniority is better everywhere even than UDOCI.  Again, the maximal seniority (and seniority zero) subspaces are the largest, which is a consequence of working with fixed local seniority.

\subsection{Intermediate Seniority}
\begin{figure*}
\includegraphics[width=\columnwidth]{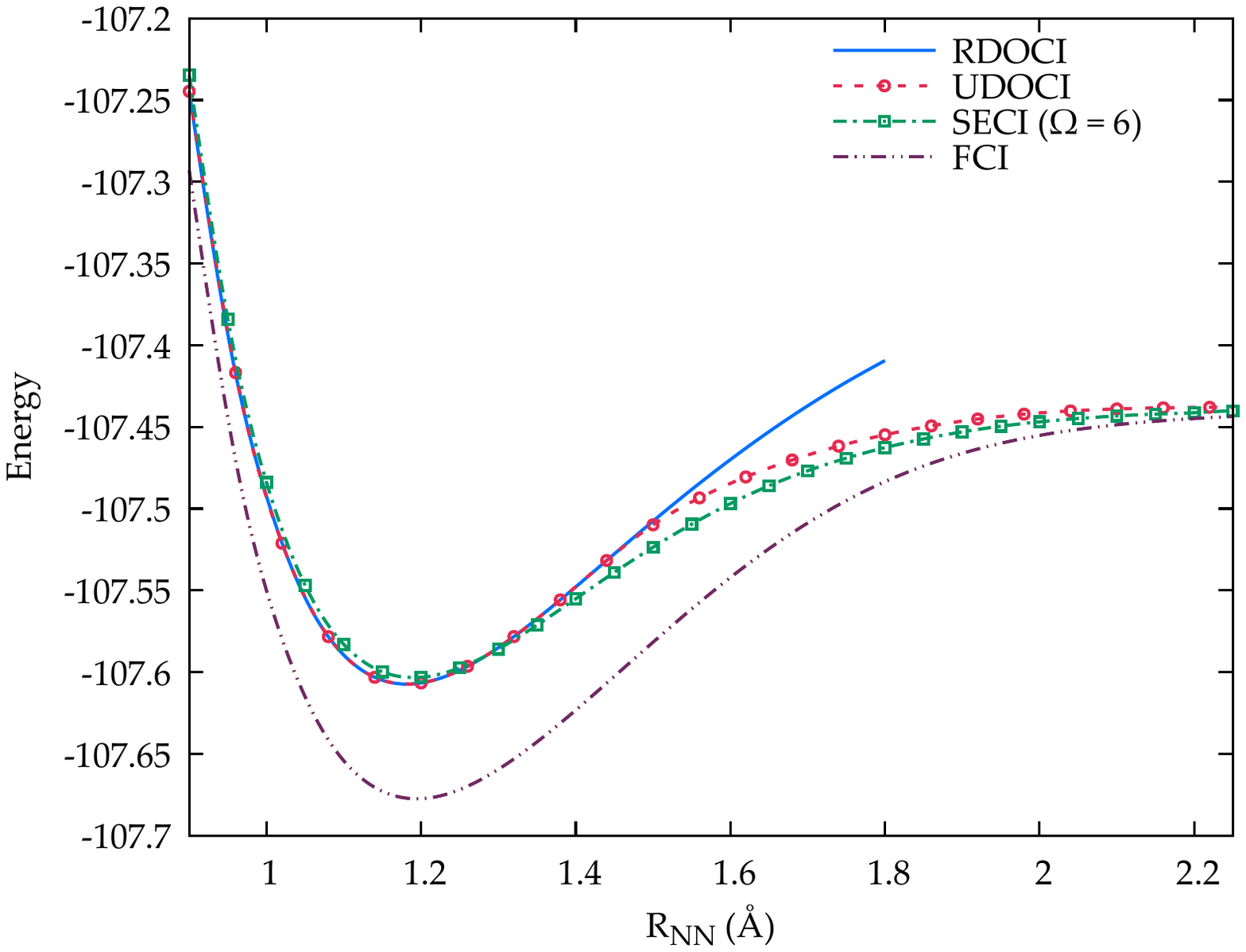}
\hfill
\includegraphics[width=\columnwidth]{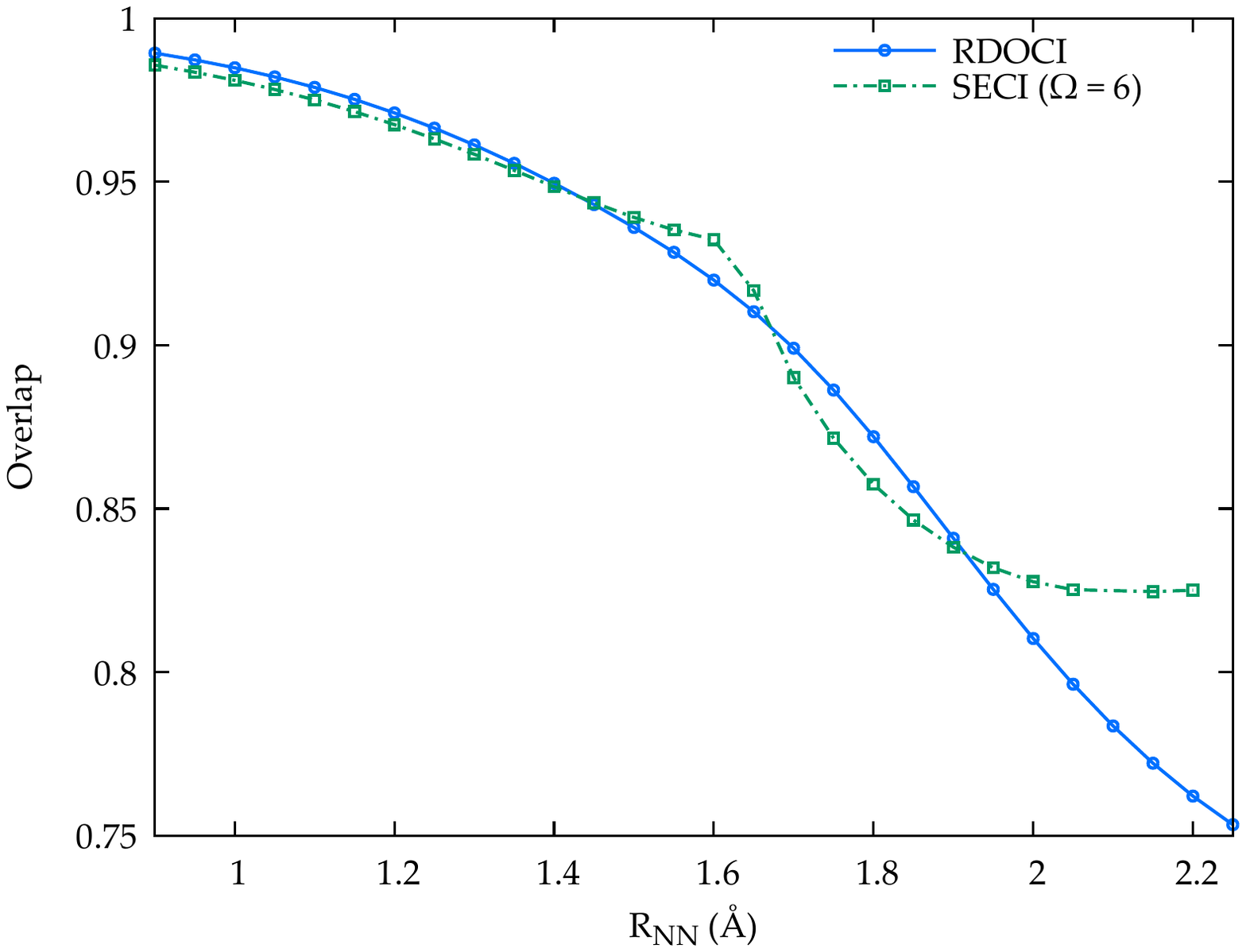}
\caption{Dissociation of N$_2$ in the STO-3G minimal basis.  Left panel: Total energies.  Right panel: Overlap with the exact wave function.
\label{Fig:N2}}
\end{figure*}

Next, we consider N$_2$ dissociation in a minimal basis (STO-3G).  We choose the $\Omega = 6$ sector, since the individual nitrogen atoms are $\Omega = 3$.  This choice suggests that at dissociation, we might expect that the 1$s$ and 2$s$ orbitals of each nitrogen atom are doubly occupied and the 2$p$ orbitals are singly occupied.  At equilibrium, it suggests doubly-occupied core orbitals but an active space of 6 singly-occupied orbitals, presumably arising from the $p$ system.  All of this implies that SECI should be quite accurate for the dissociation limit, but might not be as accurate as RDOCI near equilibrium, where we expect three doubly-occupied orbitals rather than six singly-occupied orbitals.  As shown in Fig. \ref{Fig:N2}, this is basically what we see: SECI is variationally superior to UDOCI and far better than RDOCI for large bond lengths, and is also superior in the intermediate coupling region.  Near equilibrium, RDOCI and UDOCI are identical, and are slightly better than SECI.  Both, of course, are far from FCI.  It is worth noting that SECI has only 20 determinants for seniority 6, and 120 for seniority 0.

Despite the small number of determinants, we have reasonable overlap with the exact wave function for the entire curve.  The SECI overlap saturates around 0.8, which we believe is simply a consequence of the broken spin symmetry in the SECI state.  Note that the $\Omega=6$ SECI overlap shows a kind of transition around $R_\mathrm{NN} \sim 1.6$ \AA.

We can consider a fully unrestricted version of SECI where both the pairing levels and spin levels are allowed to break spin symmetry.  In N$_2$ dissociation, these two methods differ slightly, but their differences are at the sub-milliHartree level throughout the dissociation curve.  At equilibrium, they are identical; the pairing levels, in other words, do not spontaneously break spin symmetry.  That symmetry is spontaneously broken at around 1.7~\AA, but the energetic effects are so negligible (about 0.1 milliHartree) that one might as well not bother to break the symmetry.

\begin{figure}
\includegraphics[width=\columnwidth]{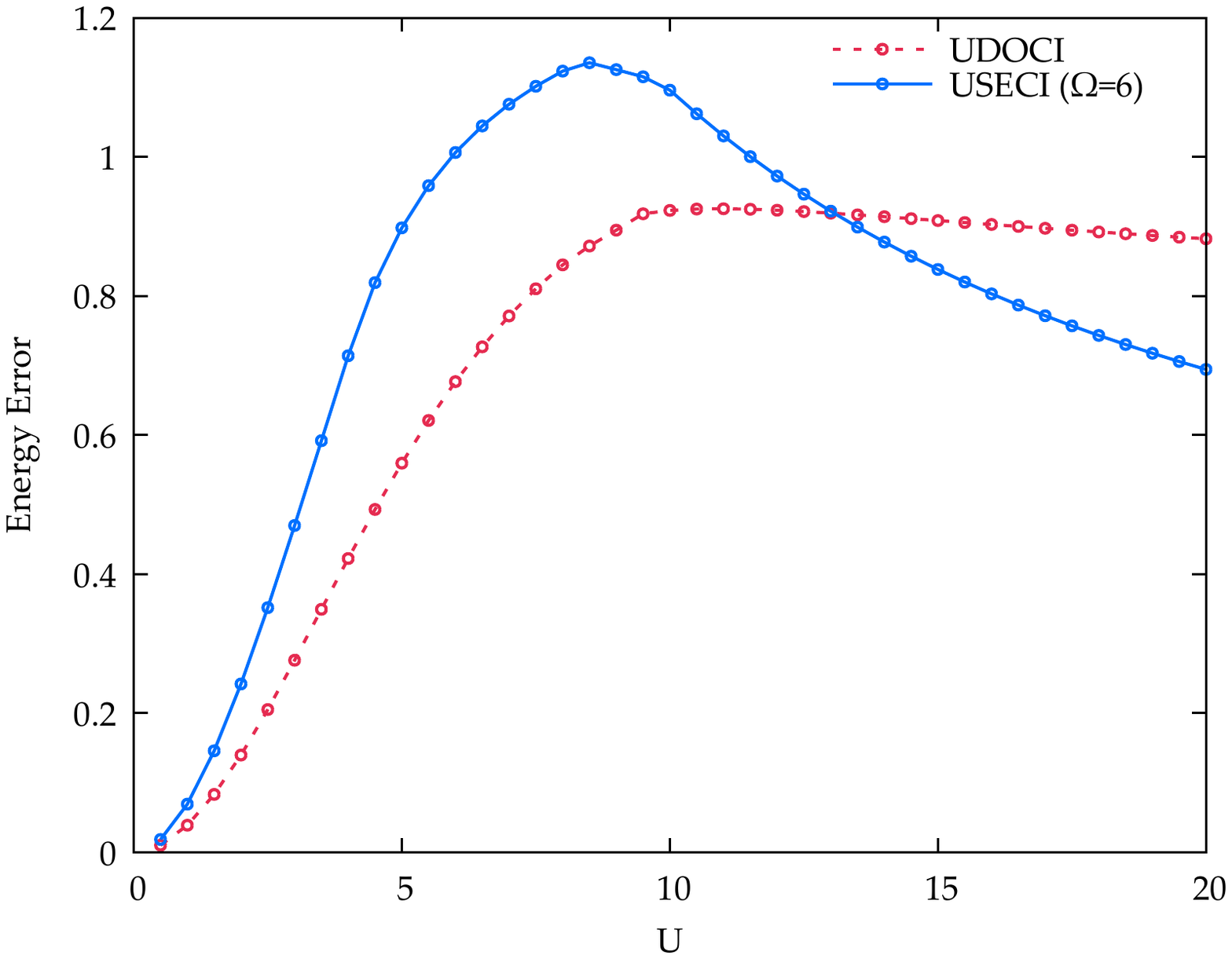}
\caption{Energy error in the 6-electron, 10-site Hubbard ring in PBC.
\label{Fig:DopedHubbard}}
\end{figure}

The same is not true in the Hubbard model.  Here, we find that restricted orbitals at maximal seniority will not do, because they give an energy independent of $U$.  While there may be a lower-energy solution that serves, we have tried instead a fully unrestricted SECI (USECI) for the doped 1D Hubbard model, as seen in Fig. \ref{Fig:DopedHubbard}. To make a fair comparison, we have compared to unrestricted DOCI (or, if one prefers, unrestricted SECI with $\Omega=0$).  For small to moderate $U$, UDOCI is the better method, while for large $U$, we again want maximal seniority.

In the on-site basis in the limit of infinitely large $U$, the ground state in the Hubbard model is a seniority eigenstate, with maximal possible seniority.  Our errors are not because we have chosen the wrong seniority sector, or the wrong orbitals (which we have optimized).  They are simply that we have chosen \textit{local} seniority, which in the lattice basis would mean that we have picked four sites in the 10-site model to be empty and six sites to be singly-occupied.  Local seniority is also a symmetry in a large $U$ Hubbard model where we have simply $H/U \to \sum_p n_{p_\uparrow} \, n_{p_\downarrow}$.  For finite $U$, we see large errors, presumably because hopping breaks seniority symmetry.

\section{A Mean-Field Approximation}
So far, everything we have discussed works in the $\mathfrak{so}(4)$ basis.  The main drawback is that we must include coupling between the pairing sector and the spin sector.  If we did not -- if we could write an approximate SECI wave function as $\ket{\Psi} \approx \left(\sum_\Gamma \xi_\Gamma \, \ket{\Xi_\Gamma}\right) \otimes \left(\sum_\Lambda  \phi_\Lambda \, \ket{\Phi_\Lambda}\right)$ so that the wave function becomes the product of a pairing part and a spin part -- we could take full advantage of the ability of methods like pair coupled cluster or Jordan-Wigner Hartree-Fock to approximate DOCI with polynomial cost.

For most of the problems we have considered here, the wave function already adopts this strictly factorized form.  It always does so for zero or maximal seniority, but it also does so for intermediate seniority whenever the pairing levels are completely filled (as in minimal basis N$_2$ with seniority 6) or completely empty (as in the 6-electron 10-site Hubbard model with seniority 6).  For this reason, we have considered a doped Hubbard model at an intermediate seniority.  Specifically, we consider the 8-site 6-electron model with seniority four. We would expect to obtain more accurate results with other seniority choices, but here we would have four spin levels with two $\uparrow$ spins and two $\downarrow$ spins and four pairing levels with one pair, the sum over spin determinants $\Gamma$ runs over six states and the sum over pairing determinants $\Lambda$ runs over four.

\begin{figure}
\includegraphics[width=\columnwidth]{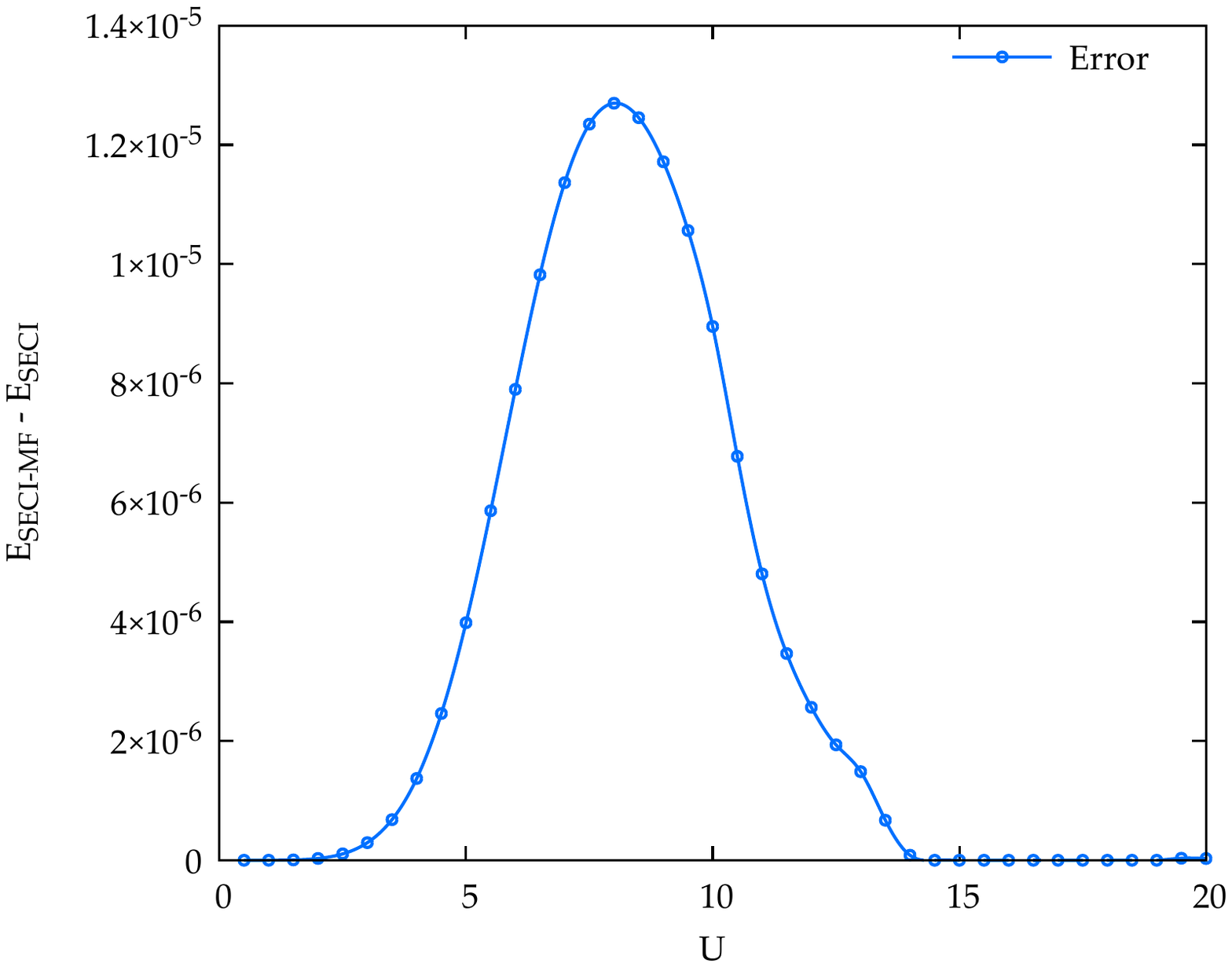}
\caption{Difference between SECI and the mean-field approximation for seniority 4 in the 6-electron 8-site Hubbard ring with PBC.
\label{Fig:MFError}}
\end{figure}

To test this idea, we do a SECI calculation, then do a singular value decomposition of the SECI coefficient $C_{\Gamma\Lambda}$ and discard all but the largest singular value, then rebuild the coefficient.  We then evaluate the energy to see how large the effect of the other singular values is.  This would form an upper bound to a true mean-field energy.  Results for the 6-electron, 8-site ring at seniority four are shown in Fig. \ref{Fig:MFError}.  Clearly, at least for this small Hubbard system, the mean-field approximation is excellent.

In practice, the approach we propose to take replaces the effective Hamiltonian of Eqn. \ref{Eqn:HSECI} with a mean-field version
\begin{align}
H_{\delta\Omega=0}^\mathrm{MF}
 &=  E_0 + \sum_\alpha \epsilon_\alpha + \frac{1}{4} \, \sum_{\alpha \ne \beta} W_{\alpha\beta} - \sum_{\mu\alpha} X_{\mu\alpha} \braket{N_\mu} \, \braket{S_\alpha^z}
\nonumber
\\
 &+ \sum_\mu \left(\epsilon_\mu + \frac{1}{2} \, \sum_\alpha W_{\mu\alpha} + \sum_\alpha X_{\mu\alpha} \braket{S_\alpha^z}\right) \, N_\mu
\nonumber
\\
 &+ \sum_{\mu\nu} L_{\mu\nu} \, P_\mu^\dagger \, P_\nu + \frac{1}{4} \, \sum_{\mu \ne \nu} W_{\mu\nu} \, N_\mu \, N_\nu
\nonumber
\\
 &+ \sum_\alpha \left(B_\alpha + \sum_{\beta \ne \alpha} X_{\beta\alpha} + \sum_\mu X_{\mu\alpha} \braket{N_\mu}\right) \, S_\alpha^z
\nonumber
\\
 &- \sum_{\alpha \ne \beta} K_{\alpha\beta}^{\uparrow\downarrow} \, S_\alpha^+ \, S_\beta^-
  + \sum_{\alpha \ne \beta} B_{\alpha\beta} \, S_\alpha^z \, S_\beta^z.
\label{Eqn:HSECIMF}
\end{align}

Our idea is simply to solve this Hamiltonian self-consistently with, for example, Jordan-Wigner Hartree-Fock \cite{Henderson2022,Chen2023,Henderson2024a,Henderson2026,Dutta2026} as the seniority zero solver for each of the mean-field--interacting parts of the Hamiltonian.  In this way, we can retain the favorable polynomial scaling of the method at the cost of needing to include a self-consistency cycle.

\section{Discussion}
Seniority is a powerful tool for organizing Hilbert space.  But several choices must be made.  Most obviously, we have to pick which orbitals to pair with which; orbital optimization, in other words, is essential.  This is not fundamentally different from the more conventional excitation level-based organization we employ, where again we must define excitation levels by first choosing a specific set of orbitals (e.g. the Hartree-Fock orbitals, or the Kohn-Sham orbitals, or the Brueckner orbitals, or the energetically optimal orbitals).  We must also decide which seniority sector or sectors to include, just as we must decide which excitation levels we want to include in our ansatz.

Unfortunately, including multiple different seniorities all at once is difficult, because while the excitation-based hierarchy yields polynomial growth with respect to excitation level starting from a single reference determinant, the size of the seniority-based truncation starts instead with a combinatorial Hilbert space at seniority zero (and at maximal seniority) and grows polynomially from there (i.e. the relative sizes of the seniority two and zero sectors differ polynomially with respect to system size).  This means that in practice we cannot thoroughly explore the seniority-based hierarchy.

Here, we have tried to show a few things.  First, and most interestingly: maximal seniority is just as relevant as zero seniority is for strong correlation.  Like zero seniority, maximal seniority is amenable to a treatment by methods such as the antisymmetrized product of one-reference orbital geminals (AP1roG), also called pair coupled cluster doubles (pCCD).   Alternatively, it may be treated by the Jordan-Wigner transformation.  We do not comment here on the degree to which pCCD reproduces maximal seniority configuration interaction.  We note, however, that the distinction between maximal and zero seniority is in some sense a matter of terminology, as we can transform one into the other by means of the Nambu transformation.

Second, we have pointed out that while maximal seniority may be useful, a system at maximal seniority is necessarily half-filled.  What we really require, presumably, is maximal seniority in some kind of active space.  Our SECI ansatz does just that.  For now we do not worry about the case that the relevant active space is not half filled.

Third, we emphasize that in SECI (and in this active space idea we have just discussed) we are conserving \textit{local} seniority: every pair of orbitals is assigned either to a spin sector where the orbitals are forced to be singly-occupied, or to a pairing sector where the orbitals are forbidden from being singly-occupied.  This dramatically reduces the size of the Hilbert space we must consider, but it also constitutes a limitation.  One may be more interested in a \textit{global} seniority eigenstate in which, for example, there are 6 singly-occupied orbitals but \textit{every} orbital is allowed to be singly occupied.  Such a wave function has vastly greater variational freedom, but the Hilbert space is also much larger and the wave function is less easy to treat computationally.

Finally, we acknowledge an important limitation.  Were the singly-occupied orbitals chosen to be closed shell, SECI would naturally factor as the product of a DOCI in the pairing space and a DOCI in the spin space.  Because of the coupling term $\sum_{\mu\alpha} X_{\mu\alpha} \, N_\mu \, S_\alpha^z$ of Eqn. \ref{Eqn:HSECI}, the exact SECI solution does not do this and has combinatorial cost (cf. Eqn. \ref{Eqn:SECIState}).  Practical calculations would require us to treat this coupling in some sort of approximate scheme.  The mean-field results for the Hubbard model are encouraging, but much more exploration is required.


\begin{acknowledgments}
This work was supported by the U.S. Department of Energy, Office of Basic Energy Sciences, under Award DE-SC0001474.  G.E.S. is a Welch Foundation Chair (C-0036).
\end{acknowledgments}

\bibliographystyle{apsrev4-2}
\bibliography{ref}

\end{document}


\title{Supplementary Material: The Seniority-Conserving Hamiltonian}

\author{Thomas M. Henderson}
\author{Guo P. Chen}
\author{Gustavo E. Scuseria}

\maketitle

Here, we wish to derive the seniority-conserving Hamiltonian in an unrestricted spinorbital basis.

\addtolength{\parskip}{\baselineskip}
The original fermionic Hamiltonian is
\begin{align}
H &=
        E_0
    +   \sum_{pq} \sum_{\sigma \in \{\uparrow,\downarrow\}} h_{pq}^\sigma \, c_{p_\sigma}^\dagger \, c_{q_\sigma}
    +   \frac{1}{2} \, \sum_{pqrs} \sum_{\sigma \in \{\uparrow,\downarrow\}} v_{pqrs}^{\sigma\sigma} \, c_{p_\sigma}^\dagger \, c_{q_\sigma}^\dagger \, c_{s_\sigma} \, c_{r_\sigma}
    +   \sum_{pqrs} v_{pqrs}^{\uparrow\downarrow} \, c_{p_\uparrow}^\dagger \, c_{q_\downarrow}^\dagger \, c_{s_\downarrow} \, c_{r_\uparrow},
\end{align}
where $v_{pqrs}$ is a two-electron integral in Dirac order and is not antisymmetrized.

The essential idea is to pair orbitals in all possible ways:
\vspace{-\baselineskip}
\begin{itemize}
\item In the one-electron pieces, this means setting $p=q$.
\item In the two-electron terms, we could set $p=q$ and $r=s$.
\item We could also set $p=r$ and $q=s$ with $p \ne q$.
\item We could also set $p=s$ and $q=r$ with $p \ne q$.
\end{itemize}
This gives us
\begin{align}
H_{\delta\Omega=0}
 &=
        E_0
    +   \sum_{p} \sum_{\sigma \in \{\uparrow,\downarrow\}} h_{pp}^\sigma \, c_{p_\sigma}^\dagger \, c_{p_\sigma}
    +   \frac{1}{2} \, \sum_{pr} \sum_{\sigma \in \{\uparrow,\downarrow\}} v_{pprr}^{\sigma\sigma} \, c_{p_\sigma}^\dagger \, c_{p_\sigma}^\dagger \, c_{r_\sigma} \, c_{r_\sigma}
    +   \sum_{pr} v_{pprr}^{\uparrow\downarrow} \, c_{p_\uparrow}^\dagger \, c_{p_\downarrow}^\dagger \, c_{r_\downarrow} \, c_{r_\uparrow}
\\
    &   +   \frac{1}{2} \, \sum_{p \ne q} \sum_{\sigma \in \{\uparrow,\downarrow\}} v_{pqpq}^{\sigma\sigma} \, c_{p_\sigma}^\dagger \, c_{q_\sigma}^\dagger \, c_{q_\sigma} \, c_{p_\sigma}
        +   \sum_{p \ne q} v_{pqpq}^{\uparrow\downarrow} \, c_{p_\uparrow}^\dagger \, c_{q_\downarrow}^\dagger \, c_{q_\downarrow} \, c_{p_\uparrow}
\nonumber
\\
    &   +   \frac{1}{2} \, \sum_{p \ne q} \sum_{\sigma \in \{\uparrow,\downarrow\}} v_{pqqp}^{\sigma\sigma} \, c_{p_\sigma}^\dagger \, c_{q_\sigma}^\dagger \, c_{p_\sigma} \, c_{q_\sigma}
        +   \sum_{p \ne q} v_{pqqp}^{\uparrow\downarrow} \, c_{p_\uparrow}^\dagger \, c_{q_\downarrow}^\dagger \, c_{p_\downarrow} \, c_{q_\uparrow}.
\nonumber
\end{align}

After using the nilpotency of $c_{p_\sigma}^\dagger$ and employing anticommutation, we arrive at
\begin{align}
H_{\delta\Omega=0}
 &=
        E_0
    +   \sum_{p} \sum_{\sigma \in \{\uparrow,\downarrow\}} h_{pp}^\sigma \, c_{p_\sigma}^\dagger \, c_{p_\sigma}
    +   \sum_{pr} v_{pprr}^{\uparrow\downarrow} \, c_{p_\uparrow}^\dagger \, c_{p_\downarrow}^\dagger \, c_{r_\downarrow} \, c_{r_\uparrow}
\\
    &   +   \frac{1}{2} \, \sum_{p \ne q} \sum_{\sigma \in \{\uparrow,\downarrow\}} v_{pqpq}^{\sigma\sigma} \, c_{p_\sigma}^\dagger \, c_{p_\sigma} \, c_{q_\sigma}^\dagger \, c_{q_\sigma}
        +   \sum_{p \ne q} v_{pqpq}^{\uparrow\downarrow} \, c_{p_\uparrow}^\dagger \, c_{p_\uparrow} \, c_{q_\downarrow}^\dagger \, c_{q_\downarrow}
\nonumber
\\
    &   -   \frac{1}{2} \, \sum_{p \ne q} \sum_{\sigma \in \{\uparrow,\downarrow\}} v_{pqqp}^{\sigma\sigma} \, c_{p_\sigma}^\dagger \, c_{p_\sigma} \, c_{q_\sigma}^\dagger \, c_{q_\sigma}
        -   \sum_{p \ne q} v_{pqqp}^{\uparrow\downarrow} \, c_{p_\uparrow}^\dagger \, c_{p_\downarrow} \, c_{q_\downarrow}^\dagger  \, c_{q_\uparrow}.
\nonumber
\end{align}

Next, we recognize
\begin{subequations}
\begin{align}
c_{p_\sigma}^\dagger \, c_{p_\sigma} &= n_{p_\sigma},
\\
c_{p_\uparrow}^\dagger \, c_{p_\downarrow}^\dagger &= P_p^\dagger,
\\
c_{r_\downarrow} \, c_{r_\uparrow} &= P_r,
\\
c_{p_\uparrow}^\dagger \, c_{p_\downarrow} &= S_p^+,
\\
c_{q_\downarrow}^\dagger  \, c_{q_\uparrow} &= S_q^-
\end{align}
\end{subequations}
for the operators, and define the Coulomb and exchange integrals
\begin{subequations}
\begin{align}
J_{pq}^{\sigma\sigma^\prime} &= v_{pqpq}^{\sigma\sigma^\prime},
\\
K_{pq}^{\sigma\sigma^\prime} &= v_{pqqp}^{\sigma\sigma^\prime}.
\end{align}
\end{subequations}
This leads us to
\begin{align}
H_{\delta\Omega=0}
 &=     E_0
    +   \sum_{p} \left(h_{pp}^\uparrow \, n_{p_\uparrow} + h_{pp}^\downarrow \, n_{p_\downarrow}\right)
    +   \sum_{pr} v_{pprr}^{\uparrow\downarrow} \, P_p^\dagger \, P_r
\\
    &   +   \frac{1}{2} \, \sum_{p \ne q} J_{pq}^{\uparrow\uparrow} \, n_{p_\uparrow} \, n_{q_\uparrow}
        +   \frac{1}{2} \, \sum_{p \ne q} J_{pq}^{\downarrow\downarrow} \, n_{p_\downarrow} \, n_{q_\downarrow}
        +   \sum_{p \ne q} J_{pq}^{\uparrow\downarrow} \, n_{p_\uparrow} \, n_{q_\downarrow}
\nonumber
\\
    &   -   \frac{1}{2} \, \sum_{p \ne q} K_{pq}^{\uparrow\uparrow} \, n_{p_\uparrow} \, n_{q_\uparrow}
        -   \frac{1}{2} \, \sum_{p \ne q} K_{pq}^{\downarrow\downarrow} \, n_{p_\downarrow} \, n_{q_\downarrow}
        -   \sum_{p \ne q} K_{pq}^{\uparrow\downarrow} \, S_p^+ \, S_q^-.
\nonumber
\end{align}

By writing
\begin{subequations}
\begin{align}
n_{p_\uparrow} &= \frac{1}{2} \, N_p + S_p^z,
\\
n_{p_\downarrow} &= \frac{1}{2} \, N_p - S_p^z,
\end{align}
\end{subequations}
we arrive at
\begin{align}
H_{\delta\Omega=0}
 &=     E_0
    +   \sum_{p} h_{pp}^\uparrow \, \left(\frac{1}{2} \, N_p + S_p^z\right)
    +   \sum_{p} h_{pp}^\downarrow \, \left(\frac{1}{2} \, N_p - S_p^z\right)
    +   \sum_{pr} v_{pprr}^{\uparrow\downarrow} \, P_p^\dagger \, P_r
\\
    &   +   \frac{1}{2} \, \sum_{p \ne q} \left(J_{pq}^{\uparrow\uparrow} - K_{pq}^{\uparrow\uparrow}\right) \, \left(\frac{1}{2} \, N_p + S_p^z\right) \, \left(\frac{1}{2} \, N_q + S_q^z\right)
        +   \frac{1}{2} \, \sum_{p \ne q} \left(J_{pq}^{\downarrow\downarrow} - K_{pq}^{\downarrow\downarrow}\right) \, \left(\frac{1}{2} \, N_p - S_p^z\right) \, \left(\frac{1}{2} \, N_q - S_q^z\right)
\nonumber
\\
    &    +   \sum_{p \ne q} J_{pq}^{\uparrow\downarrow} \,\left(\frac{1}{2} \, N_p + S_p^z\right)  \, \left(\frac{1}{2} \, N_q - S_q^z\right)
        -   \sum_{p \ne q} K_{pq}^{\uparrow\downarrow} \, S_p^+ \, S_q^-.
\nonumber
\end{align}

Grouping terms gives us
\begin{align}
H_{\delta\Omega=0}
 &=     E_0
    +   \sum_p \frac{1}{2} \, \left(h_{pp}^\uparrow + h_{pp}^\downarrow\right) \, N_p
    +   \sum_p \left(h_{pp}^\uparrow - h_{pp}^\downarrow\right) \, S_p^z
    +   \sum_{pq} v_{ppqq}^{\uparrow\downarrow} \, P_p^\dagger \, P_q
\\
    &+  \frac{1}{4} \, \sum_{p \ne q} \left(
            \frac{1}{2} \, J_{pq}^{\uparrow\uparrow} - \frac{1}{2} \, K_{pq}^{\uparrow\uparrow}
        +   \frac{1}{2} \, J_{pq}^{\downarrow\downarrow} - \frac{1}{2} \, K_{pq}^{\downarrow\downarrow}
        +   J_{pq}^{\uparrow\downarrow}\right) \, N_p \, N_q
\nonumber
\\
    &+  \sum_{p \ne q} \left(
            \frac{1}{2} \, J_{pq}^{\uparrow\uparrow} - \frac{1}{2} \, K_{pq}^{\uparrow\uparrow}
        +   \frac{1}{2} \, J_{pq}^{\downarrow\downarrow} - \frac{1}{2} \, K_{pq}^{\downarrow\downarrow}
        -   J_{pq}^{\uparrow\downarrow}\right) \, S_p^z \, S_q^z
\nonumber
\\
    &   +   \frac{1}{2} \, \sum_{p \ne q} \left(J_{pq}^{\uparrow\uparrow} - K_{pq}^{\uparrow\uparrow}\right) \, \left(\frac{1}{2} \, N_p \, S_q^z + \frac{1}{2} \, N_q \, S_p^z\right)
        -   \frac{1}{2} \, \sum_{p \ne q} \left(J_{pq}^{\downarrow\downarrow} - K_{pq}^{\downarrow\downarrow}\right) \, \left(\frac{1}{2} \, N_p \, S_q^z + \frac{1}{2} \, N_q S_p^z\right)
\nonumber
\\
    &    +   \sum_{p \ne q} J_{pq}^{\uparrow\downarrow} \,\left(-\frac{1}{2} \, N_p \, S_q^z + \frac{1}{2} \, N_q \, S_p^z\right)
        -   \sum_{p \ne q} K_{pq}^{\uparrow\downarrow} \, S_p^+ \, S_q^-.
\nonumber
\end{align}

Using symmetry of $J_{pq}^{\sigma\sigma}$ and $K_{pq}^{\sigma\sigma}$ gives us
\begin{align}
H_{\delta\Omega=0}
 &=     E_0
    +   \sum_p \frac{1}{2} \, \left(h_{pp}^\uparrow + h_{pp}^\downarrow\right) \, N_p
    +   \sum_p \left(h_{pp}^\uparrow - h_{pp}^\downarrow\right) \, S_p^z
    +   \sum_{pq} v_{ppqq}^{\uparrow\downarrow} \, P_p^\dagger \, P_q
\\
    &+  \frac{1}{4} \, \sum_{p \ne q} \left(
            \frac{1}{2} \, J_{pq}^{\uparrow\uparrow} - \frac{1}{2} \, K_{pq}^{\uparrow\uparrow}
        +   \frac{1}{2} \, J_{pq}^{\downarrow\downarrow} - \frac{1}{2} \, K_{pq}^{\downarrow\downarrow}
        +   J_{pq}^{\uparrow\downarrow}\right) \, N_p \, N_q
\nonumber
\\
    &+  \sum_{p \ne q} \left(
            \frac{1}{2} \, J_{pq}^{\uparrow\uparrow} - \frac{1}{2} \, K_{pq}^{\uparrow\uparrow}
        +   \frac{1}{2} \, J_{pq}^{\downarrow\downarrow} - \frac{1}{2} \, K_{pq}^{\downarrow\downarrow}
        -   J_{pq}^{\uparrow\downarrow}\right) \, S_p^z \, S_q^z
\nonumber
\\
    &   +   \frac{1}{2} \, \sum_{p \ne q} \left(J_{pq}^{\uparrow\uparrow} - K_{pq}^{\uparrow\uparrow}\right) \, N_p \, S_q^z
        -   \frac{1}{2} \, \sum_{p \ne q} \left(J_{pq}^{\downarrow\downarrow} - K_{pq}^{\downarrow\downarrow}\right) \, N_p \, S_q^z
        -   \frac{1}{2} \, \sum_{p \ne q} \left(J_{pq}^{\uparrow\downarrow} - J_{qp}^{\uparrow\downarrow} \right) \, N_p \, S_q^z
\nonumber
\\
    &    -   \sum_{p \ne q} K_{pq}^{\uparrow\downarrow} \, S_p^+ \, S_q^-.
\nonumber
\end{align}

We therefore recognize
\begin{align}
H_{\delta\Omega=0}
 &=     E_0
    +   \sum_p \underbrace{\frac{1}{2} \, \left(h_{pp}^\uparrow + h_{pp}^\downarrow\right)}_{\epsilon_p} \, N_p
    +   \sum_p \underbrace{\left(h_{pp}^\uparrow - h_{pp}^\downarrow\right)}_{B_p} \, S_p^z
    +   \sum_{pq} \underbrace{v_{ppqq}^{\uparrow\downarrow}}_{L_{pq}} \, P_p^\dagger \, P_q
\\
    &+  \frac{1}{4} \, \sum_{p \ne q} \underbrace{\left(
            \frac{1}{2} \, J_{pq}^{\uparrow\uparrow} - \frac{1}{2} \, K_{pq}^{\uparrow\uparrow}
        +   \frac{1}{2} \, J_{pq}^{\downarrow\downarrow} - \frac{1}{2} \, K_{pq}^{\downarrow\downarrow}
        +   J_{pq}^{\uparrow\downarrow}\right)}_{W_{pq}} \, N_p \, N_q
\nonumber
\\
    &+  \sum_{p \ne q} \underbrace{\left(
            \frac{1}{2} \, J_{pq}^{\uparrow\uparrow} - \frac{1}{2} \, K_{pq}^{\uparrow\uparrow}
        +   \frac{1}{2} \, J_{pq}^{\downarrow\downarrow} - \frac{1}{2} \, K_{pq}^{\downarrow\downarrow}
        -   J_{pq}^{\uparrow\downarrow}\right)}_{B_{pq}} \, S_p^z \, S_q^z
\nonumber
\\
    &   +   \sum_{p \ne q} \underbrace{\frac{1}{2} \, \left[
            \left(J_{pq}^{\uparrow\uparrow} - K_{pq}^{\uparrow\uparrow}\right)
        -   \left(J_{pq}^{\downarrow\downarrow} - K_{pq}^{\downarrow\downarrow}\right)
        -   \left(J_{pq}^{\uparrow\downarrow} - J_{qp}^{\uparrow\downarrow}\right)\right]}_{X_{pq}} \, N_p \, S_q^z
\nonumber
\\
    &    -   \sum_{p \ne q} K_{pq}^{\uparrow\downarrow} \, S_p^+ \, S_q^-.
\nonumber
\end{align}

We can equivalently write
\begin{equation}
X_{pq} =    \frac{1}{2} \, \left(J_{pq}^{\uparrow\uparrow} - J_{pq}^{\downarrow\downarrow}\right)
        -   \frac{1}{2} \, \left(K_{pq}^{\uparrow\uparrow} - K_{pq}^{\downarrow\downarrow}\right)
        -   \frac{1}{2} \, \left(J_{pq}^{\uparrow\downarrow} - J_{qp}^{\uparrow\downarrow}\right)
\end{equation}
which makes it clearer that $X_{pq}$ vanishes for restricted orbitals.